\documentclass{elsart}
 \usepackage{graphics}
 \usepackage{epsfig}
\usepackage{amssymb,amsmath}
\font\bba=msbm10 scaled 1080
\font\bbb=msbm8 
\font\bbc=msbm6 
\newfam\bbfam
\textfont\bbfam=\bba \scriptfont\bbfam=\bbb
\scriptscriptfont\bbfam=\bbc
\def\bb{\fam\bbfam\bba}

\def\Z{{\bb Z}}

\journal{Physica A}

\begin{document}

\begin{frontmatter}

\title{Statistical field theory for simple fluids: the collective
variables representation}
 \author[Orsay]{J.-M. Caillol}
 \address[Orsay]{Laboratoire de Physique Th\'eorique
CNRS UMR 8627, B\^at. 210,
Universit\'e de Paris-Sud,
91405 Orsay Cedex, France}
 \ead{Jean-Michel.Caillol@th.u-psud.fr}

\author[Lviv]{O. Patsahan\corauthref{cor}},
\corauth[cor]{Corresponding author.}
\ead{oksana@icmp.lviv.ua}
 \author[Lviv]{I. Mryglod}
\ead{mryglod@icmp.lviv.ua}
\address[Lviv]{Institute for Condensed Matter Physics
of the National Academy of Sciences of Ukraine,
1 Svientsitskii Str.,
79011 Lviv, Ukraine}

\begin{abstract}
An alternative representation of an exact statistical field theory
for simple fluids, based on the method of collective variables, is
presented. The results obtained are examined from the point of
another version of theory that was developed recently by
performing a Hubbard-Stratonovich transformation of the
configurational Boltzmann factor [J.-M. Caillol, Mol. Phys. 101
(2003) 1617]. The analytical expressions for the pressure and the
free energy are derived in two-loop approximation for both
versions of theory and it is shown that they are indeed
equivalent. The results yield a new type approximation within an untested
approximation scheme.
\end{abstract}

\end{frontmatter}

\section{Introduction}
\label{intro}
Functional methods in modern statistical physics represent one of
the most powerful tools for the study both of equilibrium and
dynamical properties (see, e.g. \cite{Orland,Zinn}). A great
amount of statistical field theories known in the literature are
based of the Hubbard-Stratonovich transformation
\cite{Hubbard1,Strato}, proposed in the 50ies. Nearly at the same
time another method - so-called collective variables (CV) method -
that allows in a explicit way to construct a functional
representation for many-particle interacting systems was developed
\cite{Zubar,Yuk1} and applied for the description of charged
particle systems, in particular, to the calculation of the
configurational integral of the Coulomb systems. The idea of this
method was based on: (i) the concept of collective coordinates
being appropriate for the physics of system considered (see, for
instance, \cite{Bohm-Pines}), and (ii) the integral equality
allowing to derive an exact functional representation for the
configurational Boltzmann factor. Later the CV methods was
successfully developed for the description of classical
many-particle systems \cite{Yuk-Hol} and for the phase transition
theory \cite{Yuk}.
One more functional approach, the mesoscopic field theory, was recently developed
for the study of phase transitions in ionic fluids \cite{ciach_stell}.

One of the goals of this paper is to reconsider the CV method from
the point of view of the statistical field theory  and to
compare the results obtained with those found recently by one of
us by means of the KSSHE
(Kac-Siegert-Stratonovich-Hubbard-Edwards) theory \cite{Cai-Mol}.

We formulate the method of CV in real space and consider a
one-component continuous model consisting of hard spheres
interacting through additive pair potentials. The expression
for the functional of the grand partition function is derived and
the CV action that depends upon two scalar fields - field $\rho$
connected to the number density of particles and field $\omega$
conjugate to $\rho$ - is calculated. We study correlations between
these fields as well as their relations to the density and energy
correlations of the fluid. The grand partition function of the
model is evaluated in a systematic way using a well-known method
of statistical field theory, namely the so-called loop expansion.
It consists in expanding functionally the action $\mathcal{H}$
around a saddle point, so that the lowest order (zero loop)
approximation defines the mean-field (MF) level of the theory and
the first order loop expressions correspond to the random phase
approximation (RPA). Recently \cite{Cai-Mol} this technique was
applied to the action obtained within the framework of the KSSHE
theory. In this paper we perform a two-loop expansion of the
pressure and the free energy of the homogeneous fluid which yields
a new type of approximation which we plan to test in our future
work.

The paper is organized as follows. In Section~2, starting from the
Hamiltonian, we introduce the two different functional
representations of the grand partition function based on the KSSHE
and CV methods. Here we also enter several types of statistical
field averages that are important in the further part of the
paper. In Section~3 we introduce the CV and KSSHE field
correlation functions, establish links between them as well as
their relation to the density correlation functions of the fluid.
The MF level of the KSSHE and CV field theories is formulated in
Section~4. Section~5 is devoted to the loop expansion of the grand
potential. The pressure and the free energy of the homogeneous
fluid are obtained in the two-loop approximation in Section~6. We
conclude with some final remarks in Section~7.

\section{Summary of previous works}
\label{summary}
\subsection{The model}
We consider  the case of a simple three dimensional  fluid that
consists of identical hard spheres of diameter $\sigma$ with
additional isotropic pair interactions $v(r_{ij})$ ($r_{ij}=\vert
\mathbf{x}_i -\mathbf{x}_j \vert$, $\mathbf{x}_i$ is the position
of particle "$i$"). Since $v(r)$ is arbitrary in the core, i.e.
for $r \leq \sigma$, we assume that $v(r)$ has been regularized in
such a way that  its Fourier transform $v_{q}$ is a well behaved
function of $q$ and that $v(0)$ is a finite quantity. We denote by
$\Omega$ the domain of volume $V$ occupied by the molecules of the
fluid. The fluid is at equilibrium in the grand canonical (GC)
ensemble  and we denote by $\beta=1/kT$ the inverse temperature
($k$ is the Boltzmann constant) and $\mu$ is the chemical
potential. In addition the particles are subject to an external
potential $\psi(\mathbf{x})$ and we will denote by $\nu(\mathbf{x})=\beta
(\mu-\psi(\mathbf{x}))$ the dimensionless local chemical potential. We will
stick to notations usually adopted in standard textbooks on the
theory of liquids (see e.g. \cite{Hansen}) and denote by
$w(r)=-\beta v(r)$ the negative of the dimensionless pair
interaction. Quite arbitrarily we will say that the interaction is
attractive if the Fourier transform $\widetilde{w}(q)$ is positive
for all $q$; in the converse case it  will be said repulsive.

In a given configuration
$\mathcal{C}=(N;\mathbf{x}_1 \ldots \mathbf{x}_N)$
the microscopic density of particles reads
\begin{equation}
\widehat{\rho}(\mathbf{x}\vert \mathcal{C})=
\sum_{i=1}^{N} \delta^{(3)}(\mathbf{x}-\mathbf{x}_i) \; ,
\end{equation}
and the GC partition function $\Xi\left[ \nu \right] $ can thus be written as
\begin{equation}
\label{csi}\Xi\left[ \nu \right] = \sum_{N=0}^{\infty}
\frac{1}{N!} \int_{\Omega}d1 \ldots dn \; \exp\left( -\beta
V_{\text{HS}}(\mathcal{C}) +\frac{1}{2} \left\langle
\widehat{\rho}\vert w \vert\widehat{\rho} \right\rangle +
\left\langle  \overline{\nu}\vert \widehat{\rho} \right\rangle
\right) \; ,
\end{equation}
where $i \equiv \mathbf{x}_i $ and $di\equiv d^{3}x_i$. For a given volume $V$,
$\Xi\left[ \nu \right]$ is a function of $\beta$ and a convex functional of the
local chemical potential $\nu(x)$ which we have strengthened by using a bracket.
 In eq.\ (\ref{csi}) $\exp(-\beta V_{\text{HS}}(\mathcal{C}))$ denotes the hard
sphere contribution to the Boltzmann factor in a configuration $\mathcal{C}$ and
$\overline{\nu}=\nu+\nu_S$ where
$\nu_S= - w(0)/2$ is proportional to the  self-energy of the
particles. From our hypothesis on $w(r)$,
$\nu_S$ is a finite quantity which depends however on the regularization of
the potential in the core. In the r.h.s of eq.\ (\ref{csi}) we have also
introduced Dirac's brac-kets notations
\begin{subequations}
\begin{eqnarray}
\left\langle  \overline{\nu}\vert \widehat{\rho} \right\rangle
 &\equiv&  \int_{\Omega} d1 \; \overline{\nu}(1)\widehat{\rho}(1) \\
\left< \widehat{\rho} \vert w \vert\widehat{\rho} \right> & \equiv &
\int_{\Omega} d1 d2\;
\widehat{\rho}(1\vert \mathcal{C})
 w(1,2)  \widehat{\rho}(2\vert \mathcal{C}) \; .
\end{eqnarray}
\end{subequations}
Previous work have shown that $\Xi\left[ \nu \right]$ can be rewritten as a functional integral.
Two  such equivalent representations are reviewed below.

\subsection{The KSSHE representation}
As it is well known the GC partition function $\Xi\left[ \nu
\right]$ can be re-expressed as a functional integral by
performing the KSSHE transformation
\cite{Kac,Siegert,Strato,Hubbard1,Hubbard2,Edwards} of the
Boltzmann's factor. Under this transformation $\Xi\left[ \nu
\right]$ can be rewritten as the GC partition function of a fluid
of bare hard spheres in the presence of a  random Gaussian field $
\varphi$  with a covariance given by the pair potential
\cite{Wiegel,Wegner,Cai-Mol,Cai-JSP}. More precisely one has
\begin{itemize}
\item \textit{ i) in the attractive  case ($\widetilde{w}(q)>0$)}
\begin{eqnarray}
\label{attractive}
\Xi\left[ \nu \right] &=&
\mathcal{N}_{w}^{-1} \int \mathcal{D} \varphi \;
\exp \left( -\frac{1}{2}
\left\langle \varphi \vert w^{-1} \vert \varphi \right\rangle \right)
\Xi_{\text{HS}}\left[ \overline{\nu} + \varphi\right]
 \nonumber \\
 &\equiv &
 \left\langle \Xi_{\text{HS}}\left[ \overline{\nu} + \varphi\right]
 \right\rangle_{w} \; ,
\end{eqnarray}
\item \textit{ ii) in the repulsive  case ($\widetilde{w}(q)<0$)}
\begin{eqnarray}
\label{repulsive}
\Xi\left[ \nu \right] &=&
\mathcal{N}_{(-w)}^{-1} \int \mathcal{D} \varphi \;
\exp \left( \frac{1}{2}
\left\langle \varphi \vert w^{-1} \vert \varphi \right\rangle \right)
\Xi_{\text{HS}}\left[ \overline{\nu} + i \varphi\right]
 \nonumber \\
 &\equiv &
 \left\langle \Xi_{\text{HS}}\left[ \overline{\nu} + i \varphi\right]
 \right\rangle_{(-w)} \; ,
\end{eqnarray}

\end{itemize}
where, in both cases, $\varphi$ is a real random field and
$\Xi_{\text{HS}}$ denotes the GC partition function of bare hard
spheres. $\Xi$ can thus be written as a Gaussian average
$\left\langle \ldots \right\rangle_{w}$ of covariance $w$ and we
have noted by $\mathcal{N}_{w}$ the normalization constant
\begin{equation}
 \label{normaw}
 \mathcal{N}_{w}= \int \mathcal{D} \varphi \;  \exp \left( -\frac{1}{2}
\left\langle \varphi \vert w^{-1} \vert \varphi \right\rangle \right) \; .
\end{equation}
The functional integrals which enter
eqs.~(\ref{attractive}),~(\ref{repulsive}) and ~(\ref{normaw}) can
be given a precise meaning in the case where the domain $\Omega$
is a cube of side $L$ with periodic boundary conditions (PBC)
which will be implicitly assumed henceforth. More details are
given  in Appendix A.

In the repulsive case, the hard core part of the interaction is not compulsory for
the existence of a thermodynamic limit \cite{Ruelle} and the reference system can be
chosen as the ideal gas \cite{Efimov,Lambert}.

Eqs~(\ref{attractive}) and ~(\ref{repulsive}) are easily generalized to the case of molecular
fluids or mixtures, for instance a charged hard spheres mixture (primitive model)\cite{Cai-JSP,Cai-Mol1}.

When the pair interaction $w$ is neither attractive nor repulsive it is necessary to introduce two
real scalar fields if some rigor is aimed at \cite{Cai-Mol}. Alternatively, eq.~(\ref{attractive})
can be considered to hold in any case having in mind that $\varphi$ can be a complex scalar field.
Therefore we shall write formally in all cases
\begin{equation}
\label{csiKSSHE}
\Xi\left[ \nu \right]=\mathcal{N}_{w}^{-1} \int \mathcal{D} \varphi \;
\exp \left( - \mathcal{H}_{\text{K}} \left[\nu, \varphi \right] \right) \; ,
\end{equation}
where the action of the KSSHE field theory reads as
\begin{equation}
\label{action-K}
\mathcal{H}_K \left[\nu, \varphi \right]= \frac{1}{2}\left\langle \varphi \vert w^{-1} \vert \varphi \right\rangle -
\ln \Xi_{\text{HS}}\left[ \overline{\nu} +  \varphi\right] \; .
\end{equation}

\subsection{The CV representation}
We introduce now briefly the CV representation of $\Xi\left[ \nu
\right]$ and refer the reader to the literature for a more
detailed presentation \cite{Yuk,Yuk1,Yuk-Hol,Yuk2,Yuk3}. The
starting point is the identity
\begin{eqnarray}
\label{a}
\exp \left(
\frac{1}{2}\left\langle \widehat{\rho}\vert w \vert \widehat{\rho}\right\rangle
\right)& =& \int \mathcal{D} \rho \;
 \delta_{\mathcal{F}}\left[ \rho -\widehat{\rho} \right]
 \exp \left(
\frac{1}{2}\left\langle \rho \vert w \vert \rho \right\rangle
\right)
 \; \nonumber \\
&=& \int \mathcal{D} \rho  \mathcal{D} \omega \;
 \exp \left(  \frac{1}{2}\left\langle \rho \vert w \vert \rho \right\rangle
 +i \left\langle \omega \vert \left\lbrace
 \rho - \widehat{\rho}
  \right\rbrace \right\rangle
 \right),
\end{eqnarray}
where we have made use of the functional "delta function"
\cite{Orland} $\delta_{\mathcal{F}}\left[ \rho \right]$ defined in
eq.~(\ref{deltaF}) in \mbox{appendix A}. Inserting eq.~(\ref{a})
in the expression~(\ref{csi}) of the GC partition function one
finds
\begin{equation}
\label{csiCV} \Xi\left[ \nu \right]= \int \mathcal{D} \rho
 \mathcal{D} \omega \;
\exp \left( - \mathcal{H}_{\text{CV}}\left[\nu, \rho, \omega \right] \right) \; ,
\end{equation}
where the action of the CV field theory reads as
\begin{equation}
\label{actionCV}
\mathcal{H}_{\text{CV}} \left[\nu, \rho, \omega \right]= -\frac{1}{2}
\left\langle \rho \vert w \vert \rho \right\rangle  - i \left\langle \omega \vert \rho\right\rangle -
\ln \Xi_{\text{HS}}\left[ \overline{\nu} - i   \omega \right] \; .
\end{equation}
We stress that $\omega$ and $\rho$ are two real scalar fields  and
that eqs.~(\ref{csiCV}) and~(\ref{actionCV}) are valid for
repulsive, attractive as well as  arbitrary pair interactions.
Moreover, with the clever normalization of Wegner \cite{Wegner}
for the functional measures there are no unspecified
multiplicative constant involved in eq.~(\ref{csiCV}) (see
Appendix A for more details).

The CV transformation is clearly  more general than the KSSHE
transformation since it  can be used for a pair interaction
$w(1,2)$ which does not possess an inverse and is easily
generalized for n-body interactions with $n>2$. The equivalence of
the CV and KSSHE representations~(\ref{csiKSSHE})
and~(\ref{csiCV}) of $\Xi\left[\nu \right]$ is readily established
in the repulsive case ( $w<0$) by making use of the properties of
Gaussian integrals (cf. eq.~(\ref{Gauss}) of Appendix A). In the
attractive or in the general case we did not find a convincing way
to obtain one formula from the other.

\subsection{Statistical average}
In the sequel it will be important to distinguish carefully, besides the usual GC average
$<\mathcal{A}(\mathcal{C})>_{\text{GC}}$ of a dynamic variable $\mathcal{A}(\mathcal{C})$,
between several types of statistical field averages.
Firstly  Gaussian averages of covariance $w$
\begin{equation}
\label{moyw}
\left\langle \mathcal{A}\left[\varphi \right] \right\rangle_{w}=
\mathcal{N}_{w}^{-1} \int \mathcal{D} \varphi \;
\exp \left(-\frac{1}{2}
\left\langle \varphi \vert w^{-1} \vert \varphi \right\rangle \right)
\mathcal{A}\left[\varphi \right],
\end{equation}
where $\mathcal{A}\left[ \varphi\right]$ is some functional of the
KSSHE field $\varphi$ and $\mathcal{N}_{w}$ has been defined in
eq.~(\ref{normaw}); secondly the KSSHE averages defined as
\begin{equation}
\label{moyK}
\left\langle \mathcal{A}\left[\varphi \right] \right\rangle_{\text{K}}= \Xi
\left[\nu \right]^{-1} \; \int \mathcal{D} \varphi \; \mathcal{A}\left[\varphi \right]
\exp \left( - \mathcal{H}_{\text{K}} \left[\nu, \varphi \right] \right),
\end{equation}
and thirdly the CV averages defined in a similar way as
\begin{equation}
\label{moyCV} \left\langle \mathcal{A}\left[\rho, \omega \right]
\right\rangle_{\text{CV}}=\Xi\left[\nu \right]^{-1} \; \int
\mathcal{D} \rho
 \mathcal{D} \omega \; \; \mathcal{A}\left[\rho, \omega \right]
\exp \left( - \mathcal{H}_{\text{CV}} \left[\nu, \rho, \omega \right] \right),
\end{equation}
where $\mathcal{A}\left[ \rho, \omega \right]$ is a functional of the two CV fields $\rho$ an $\omega $.

\section{Correlation functions}
\label{Corre}
Since all thermodynamic quantities of the fluid can
be expressed in terms of the  GC density correlation functions
$G^{(n)}$ \cite{Hansen} it is important to relate them to the
field correlation functions in the CV and the KSSHE
representations; this is subject of  section~\ref{Corre}.

\subsection{Density correlations}
\label{CorreGC}
The ordinary and truncated (or connected)  density correlation
functions of the fluid will be defined in this paper as
\cite{Hansen,Stell1,Stell2}
\begin{eqnarray}
\label{defcorre}
G^{(n)}[\nu](1, \ldots, n) &=&\left< \prod_{1=1}^{n} \widehat{\rho}
    (\mathbf{x}_{i}  \vert \mathcal{C}) \right>_{GC}
=\frac{1}{\Xi[\nu]}\frac{\delta^{n} \;\Xi[\nu]}
{\delta \nu(1) \ldots \delta \nu(n)}           \; ,\nonumber \\
G^{(n), T}[\nu](1, \ldots, n) &=&  \frac{\delta^{n} \log \Xi[\nu]}
{\delta \nu(1) \ldots \delta \nu(n)} \; .
\end{eqnarray}
Our notation emphasizes the fact that the  $G^{(n)}$
(connected and not connected) are functionals of the local chemical potential
 $\nu(x)$ and
functions of the coordinates $(1,\ldots, n) \equiv (\mathbf{x}_{1},\ldots,
\mathbf{x}_{n})$.

We know from the theory of liquids that
\cite{Stell1,Stell2}
\begin{equation}
G^{(n), T}[\nu](1,\ldots,n)= G^{(n)}[\nu]( 1,\ldots,n)
- \sum \prod_{m<n}G^{(m), T} [\nu](i_{1},\ldots,i_{m})  \; ,
\end{equation}
where the sum of products is carried out over all possible partitions of
the set $(1,\ldots,n)$ into subsets of cardinal $m<n$. Of course $\rho[\nu](x) \equiv
G^{(n=1)}[\nu](x)=G^{(n=1), T}[\nu](x)$ is the local density of the fluid.

It follows from the definition~(\ref{defcorre}) of the $G^{(n)}[\nu](1, \ldots, n)$
that they can be reexpressed as  KSSHE or  CV statistical averages, i.e.
\begin{subequations}
\begin{eqnarray}
G^{(n)}[\nu](1, \ldots, n)&=&\left\langle
G^{(n)}_{\text{HS}}[\overline{\nu} + \varphi](1, \ldots,
n)\right\rangle_{\text{K}} \; ,  \\ G^{(n)}[\nu](1, \ldots,
n)&=&\left\langle G^{(n)}_{\text{HS}}[\overline{\nu}  -
i\omega](1, \ldots, n)\right\rangle_{\text{CV}} \; .
\end{eqnarray}
\end{subequations}
Although enlightening these relations are not very useful except
for the special  case $n=1$ which reads explicitly as
\begin{subequations}
\begin{eqnarray}
\rho\left[\nu \right](\mathbf{x})&=& \left\langle
\rho_{\text{HS}}[\overline{\nu} +
\varphi](\mathbf{x})\right\rangle_{\text{K}} \; , \\ \rho\left[\nu
\right](\mathbf{x})&=& \left\langle \rho_{\text{HS}}[\overline{\nu}-
i\omega ](\mathbf{x})\right\rangle_{\text{CV}} \; ,
\end{eqnarray}
\end{subequations}
where $\rho_{\text{HS}}[\xi](\mathbf{x})$ is the local density of the hard
sphere fluid with the local chemical potential $\xi(\mathbf{x})$.

\subsection{KSSHE field correlations}
\label{CorreK}
Let us introduce the modified partition function
\begin{equation}
\Xi^{1}\left[ \nu, J \right]=\mathcal{N}_{w}^{-1} \int \mathcal{D} \varphi \;
\exp \left( - \mathcal{H}_{\text{K}} \left[\nu, \varphi \right] +
 \left\langle J \vert \varphi\right\rangle
 \right) \; ,
\end{equation}
where $J$ is a real scalar field. Clearly $\Xi^{1}\left[ \nu, J \right]$ is the generator of field
correlation functions and a standard result of
statistical field theory yields \cite{Zinn}
\begin{eqnarray}
\label{defcorreK} G^{(n)}_{\varphi}[\nu](1, \ldots, n) &=&\left<
\prod_{i=1}^{n} \varphi \left(\mathbf{x}_{i}\right)
\right>_{\text{K}} \;
=\frac{1}{\Xi[\nu]} \left.
\frac{\delta^{n} \;\Xi^1[\nu,J]} {\delta J(1) \ldots \delta J(n)}
\right \vert_{J=0}          \; ,\nonumber \\ G^{(n), T}_{\varphi}
[\nu](1, \ldots, n) &=& \left. \frac{\delta^{n} \; \log
\Xi^1[\nu,J]} {\delta J(1) \ldots \delta J(n)}\right \vert_{J=0}
\; .
\end{eqnarray}
Moreover one has  \cite{Zinn}
\begin{equation}
G^{(n), T}_{\varphi} (1,\ldots,n)= G^{(n)}_{\varphi}( 1,\ldots,n)
- \sum \prod_{m<n}G^{(m), T}_{\varphi}(i_{1},\ldots,i_{m})  \; .
\end{equation}
The relations between the $G^{(n), T}_{\varphi}$ and the truncated density correlation functions
$G^{(n), T}$ have been established elsewhere by one of us \cite{Cai-Mol,Cai-JSP,Cai-Mol1}.
We summarize them below for future reference.
\begin{subequations}
\label{dens-K}
\begin{eqnarray}
\label{dens-K1} \rho\left[ \nu \right](1)&=& w^{-1}(1,1^{'})
\left\langle \varphi (1^{'}) \right\rangle_{\text{K}} \; , \\
\label{dens-K2}
 G^{(2), T}\left[ \nu \right] (1,2)     &=& -w^{-1}(1,2)
 +w^{-1}(1,1^{'})w^{-1}(2,2^{'}) \times \nonumber \\
&\times & G^{(2), T}_{\varphi}\left[ \nu \right] (1^{'},2^{'}) \;
, \\
 \label{dens-Kn}
 G^{(n), T}\left[ \nu \right] (1,\ldots,n)&=& w^{-1}(1,1^{'})
 \ldots w^{-1}(n,n^{'}) \times \nonumber \\
 &\times &
 G^{(n), T}_{\varphi}\left[ \nu \right] (1',\ldots,n') \qquad
 \text{ for } n\geq 3 \; ,
\end{eqnarray}
\end{subequations}
where  we have adopted Einstein's convention, i.e. a space integration  of position
variables labled by the same dummy indices  over the domain $\Omega$ is meant.

\subsection{\label{CorreCV}CV field correlations}
In this section we study the correlation functions of the fields $\rho$ and $\omega$ in
the CV representation. We thus define
\begin{eqnarray}
\label{G-CV} G^{(n)}_{\rho}[\nu](1, \ldots, n) &=&\left<
\prod_{1=1}^{n} \rho \left(\mathbf{x}_{i}\right)
\right>_{\text{CV}} \; , \nonumber \\ G^{(n)}_{\omega}[\nu](1,
\ldots, n) &=&\left< \prod_{1=1}^{n} \omega
\left(\mathbf{x}_{i}\right) \right>_{\text{CV}} \; ,
\end{eqnarray}
and their connected parts
\begin{eqnarray}
G^{(n), T}_{\rho} (1,\ldots,n)&= & G^{(n)}_{\rho}( 1,\ldots,n) -
\sum \prod_{m<n}G^{(m), T}_{\rho}(i_{1},\ldots,i_{m}) \; ,
\nonumber \\ G^{(n), T}_{\omega} (1,\ldots,n)&= &
G^{(n)}_{\omega}( 1,\ldots,n) - \sum \prod_{m<n}G^{(m),
T}_{\omega}(i_{1},\ldots,i_{m}) \; . \;
\end{eqnarray}
\subsubsection{Correlation functions $G^{(n)}_{\rho}$}
\label{gro}
Let us define the modified GC partition function
\begin{equation}
\label{Xi2}
\Xi^2\left[\nu,J \right]=
 \int \mathcal{D} \rho
 \mathcal{D} \omega \;
\exp \left( - \mathcal{H}_{\text{CV}}\left[\nu, \rho, \omega \right]+\left\langle  J \vert \rho\right\rangle \right) \; ,
\end{equation}
where $J$ is a real scalar field. $\Xi^2\left[\nu,J \right]$ is clearly the generator of the $G^{(n)}_{\rho}$
and we thus have
\begin{eqnarray}
\label{defcorreCVrho}
G^{(n)}_{\rho}[\nu](1, \ldots, n) &=& \frac{1}{\Xi[\nu]} \left. \frac{\delta^{n} \;\Xi^2[\nu,J]}
{\delta J(1) \ldots \delta J(n)} \right \vert_{J=0}          \; ,\nonumber \\
G^{(n), T}_{\rho} [\nu](1, \ldots, n) &=& \left.  \frac{\delta^{n} \log \Xi^2[\nu,J]}
{\delta J(1) \ldots \delta J(n)}  \right \vert_{J=0}\; .
\end{eqnarray}
The simplest way to obtain the relations between the
$G^{(n)}_{\rho}$ and the density correlation functions is to start
from the definition~(\ref{defcorre}) of $G^{(n)}$. One has
\begin{eqnarray}
G^{(n)}(1, \ldots, n)&=& \frac{1}{\Xi[\nu]}\frac{\delta^{n}
\;\Xi[\nu]} {\delta \nu(1) \ldots \delta \nu(n)}           \;
\nonumber \\ &=&\frac{1}{\Xi[\nu]}
 \int \mathcal{D} \rho
 \mathcal{D} \omega \;
 \exp \left(\frac{1}{2} \left\langle\rho \vert w\vert \rho \right\rangle
 +i \left\langle \omega\vert \rho \right\rangle \right)
 \frac{\delta^{n} \;
 \Xi_{\text{HS}}[\overline{\nu} -i \omega]}
{\delta \nu(1) \ldots \delta \nu(n)} \nonumber
 \\
&=&\frac{1}{\Xi[\nu]}
 \int \mathcal{D} \rho
 \mathcal{D} \omega \;
 \exp \left(\frac{1}{2} \left\langle\rho \vert w\vert \rho \right\rangle
 +i \left\langle \omega\vert \rho \right\rangle \right) \times \nonumber \\
 &\times& (i)^n
 \frac{\delta^{n} \;
 \Xi_{\text{HS}}[\overline{\nu} -i \omega]}
{\delta \omega(1) \ldots \delta \omega(n)} \nonumber \; .
\end{eqnarray}
Performing now $n$ integral by parts yields
\begin{eqnarray}
G^{(n)}(1, \ldots, n)&=&\frac{(-i)^n}{\Xi[\nu]}
\int \mathcal{D} \rho \mathcal{D} \omega \;
 \exp \left(\frac{1}{2} \left\langle\rho \vert w\vert \rho \right\rangle
 + \ln \Xi_{\text{HS}}[\overline{\nu} -i \omega] \right)\times  \nonumber \\
 &\times&
 \frac{\delta^{n} \;
 \exp\left(i\left\langle \omega \vert \rho \right\rangle \right) }
{\delta \omega(1) \ldots \delta \omega(n)}
 =\left< \prod_{1=1}^{n} \rho \left(i\right) \right>_{\text{CV}}
  \; . \nonumber
\end{eqnarray}
We have thus proved the expected result
\begin{equation}
G^{(n)}\left[\nu \right] (1, \ldots, n)=G^{(n)}_{\rho}\left[\nu \right] (1, \ldots, n)
\; ,
\end{equation}
which implies in turn that
\begin{equation}
\label{dens-CV-rho}G^{(n), T}\left[\nu \right] (1, \ldots, n)=G^{(n),T}_{\rho}\left[\nu \right] (1, \ldots, n)
\; .
\end{equation}
\subsubsection{Correlation functions $G^{(n)}_{\omega}$}
\label{Gnomega}
Let us define  the modified GC partition function
\begin{equation}
\label{Xi3}
\Xi^3\left[\nu,J \right]=
 \int \mathcal{D} \rho
 \mathcal{D} \omega \;
\exp \left( - \mathcal{H}_{\text{CV}} \left[\nu, \rho, \omega
\right]+\left\langle  J \vert \omega \right\rangle \right) \; ,
\end{equation}
where $J$ is a real scalar field. $\Xi^3\left[\nu,J \right]$ is
the generator of the $G^{(n)}_{\omega}$ and we thus have
\begin{eqnarray}
\label{defcorreCVomega} G^{(n)}_{\omega}[\nu](1, \ldots, n) &=&
\frac{1}{\Xi[\nu]} \left. \frac{\delta^{n} \;\Xi^3[\nu,J]} {\delta
J(1) \ldots \delta J(n)} \right \vert_{J=0}          \; ,\nonumber
\\ G^{(n), T}_{\omega} [\nu](1, \ldots, n) &=& \left.
\frac{\delta^{n} \log \Xi^3[\nu,J]} {\delta J(1) \ldots \delta
J(n)}  \right \vert_{J=0}\; .
\end{eqnarray}
In order to relate $G^{(n)}_{\omega}$ to $G^{(n)}$ we perform the
change of variable $\rho \rightarrow \rho + i J$ in
eq.~(\ref{Xi3}). The functional Jacobian of the transformation is
of course equal to unity and one obtains the relation
\begin{equation}
\label{jojo} \ln \Xi^3\left[\nu,J \right]= -\frac{1}{2}
\left\langle J\vert w \vert J  \right\rangle + \ln\Xi^2\left[\nu,
i w \star J \right] \;,
\end{equation}
where the star $\star$ means a space convolution and $\Xi^2$ was
defined at eq.~(\ref{Xi2}). The idea is to perform now $n$
successive functional derivatives   of both sides of
eq.~(\ref{jojo}) with respect to $J$. Since it follows from the
expression~(\ref{defcorreCVrho}) of $G^{(n), T}_{\rho} $ that
\begin{eqnarray}
\left.  \frac{\delta^{n} \log \Xi^2[\nu, i w \star J]}
{\delta J(1) \ldots \delta J(n)}  \right \vert_{J=0}&=&
i^n w(1,1^{'}) \ldots w(n,n^{'})\times \nonumber \\
&\times&  G^{(n), T}_{\rho} [\nu](1^{'}, \ldots, n^{'}) \; ,
\end{eqnarray}
one readily obtains
\begin{subequations}
\label{dens-CV}
\begin{eqnarray}
\label{dens-CV1} \left\langle \omega (1)
\right\rangle_{\text{CV}}&=& i \; w(1,1^{'}) \left\langle \rho
(1^{'}) \right\rangle_{\text{CV}} \; , \\ \label{dens-CV2}
 G^{(2), T}_{\omega}\left[ \nu \right] (1,2)     &=&- w(1,2)  \nonumber \\
 & - &w(1,1^{'})w(2,2^{'})
 G^{(2), T}_{\rho}\left[ \nu \right] (1^{'},2^{'}) \; , \\
 \label{dens-CVn}
 G^{(n), T}_{\omega}\left[ \nu \right] (1,\ldots,n)&=&  i^n \;  w(1,1^{'})
 \ldots w(n,n^{'}) \times \nonumber \\
 &\times &
 G^{(n), T}_{\rho}\left[ \nu \right] (1',\ldots,n')  \qquad \text{ for } n\geq 3 \; .
\end{eqnarray}
\end{subequations}
A comparison between eqs~(\ref{dens-K}), ~(\ref{dens-CV-rho}),
and ~(\ref{dens-CV}) will show that the CV and KSSHE correlation
functions are simply related since we have, as expected
\begin{equation}
  G^{(n), T}_{\varphi}\left[ \nu \right] (1,\ldots,n)=
  (-i)^n \; G^{(n), T}_{\omega}\left[ \nu \right] (1,\ldots,n) \;
  \qquad \text{ for all }  n \; .
\end{equation}
\section{Mean Field theory}
\label{MF}
\subsection{\label{MF-KSSHE}KSSHE representation}
We summarize previous work on the mean-field (MF) or saddle point
approximation of the KSSHE theory \cite{Cai-Mol,Cai-JSP}. At the
MF level one has \cite{Zinn}
\begin{equation}
\Xi_{\text{MF}}\left[\nu \right]=\exp \left(- \mathcal{H}_{\text{K}}\left[
\nu, \varphi_{0} \right]  \right) \; ,
\end{equation}
where, for $\varphi=\varphi_{0}$, the action is stationary, i.e.
\begin{equation}
\label{statio-K}\left. \frac{\delta \; \mathcal{H}_{\text{K}}\left[
\nu, \varphi \right]}{\delta \varphi}\right \vert_{\varphi_{0}}=0 \; .
\end{equation}
Replacing the KSSHE action by its expression~(\ref{action-K}) in eq.~(\ref{statio-K})
leads to an  implicit equation for $\varphi_{0}$:
\begin{equation}
\label{statio2-K}
\varphi_{0}(1)=w(1,1^{'}) \; \rho_{\text{HS}}\left[\overline{\nu} +
\varphi_{0}\right]( 1^{'}) \; ,
\end{equation}
which reduces to
\begin{equation}
\label{statio2-K-hom}
\varphi_{0}=\widetilde{w}(0)\; \rho_{\text{HS}}\left[\overline{\nu} +
\varphi_{0}\right] \;
\end{equation}
for a homogeneous system.
It follows from the stationary condition~(\ref{statio-K}) that the MF density is given by
\begin{equation}
\label{ro-MF}
\rho_{\text{MF}}\left[\nu \right] (1)= \frac{\delta \ln \Xi_{\text{MF}}\left[\nu \right]}{\delta \nu(1)}=
\rho_{\text{HS}} \left[ \overline{\nu} +
\varphi_{0} \right](1) \; ,
\end{equation}
and that the MF grand potential reads
\begin{equation}
\label{MF-gpot}
\ln \Xi_{\text{MF}}\left[\nu \right]=
\ln \Xi_{\text{HS}}\left[\overline{\nu} +
\varphi_{0}\right] -  \frac{1}{2}\left\langle \rho_{\text{MF}}\vert w \vert\rho_{\text{MF}}\right\rangle \; .
\end{equation}
Moreover, the MF Kohn-Scham free energy defined as the Legendre transform
\begin{equation}
\beta \mathcal{A}_{\text{MF}}\left[\rho \right]  =\sup_{\nu}\left\lbrace
 \left\langle \rho \vert \nu\right\rangle -\ln \Xi_{\text{MF}}\left[\nu \right]
 \right\rbrace
\end{equation}
is found to be
\begin{equation}
\label{MF-A}
\beta \mathcal{A}_{\text{MF}}\left[\rho \right]  =
\beta \mathcal{A}_{\text{HS}}\left[\rho \right]
-\frac{1}{2}\left\langle\rho \vert w \vert \rho \right\rangle +\frac{1}{2} \int_{\Omega} dx \; w(0) \rho(x) \; .
\end{equation}
It can be shown \cite{Cai-Mol} that $\mathcal{A}_{\text{MF}}\left[\rho \right]$
constitutes a rigorous upper bound for the exact free energy
$\mathcal{A}\left[\rho \right]$  if the interaction is attractive
($\widetilde{w}(q)>0$) and a lower bound in the converse case ($\widetilde{w}(q)<0$).

Finally, the pair correlation  and vertex functions at the
zero-loop order which are defined respectively as
\begin{subequations}
\begin{eqnarray}
G_{\text{MF}}^{(2), T}\left[\nu \right](1,2)&=& \frac{\delta^{2} \ln \Xi_{\text{\text{MF}}}\left[ \nu\right]}
{\delta \nu(1) \; \delta \nu(2)} \; , \\
C_{\text{MF}}^{(2)}\left[\rho \right] (1,2)&=& -\frac{\delta^{2}
\beta \mathcal{A}_{\text{\text{MF}}}\left[\rho\right]} {\delta
\rho(1) \; \delta \rho(2)} \; ,
\end{eqnarray}
\end{subequations}
are given by
\begin{subequations}
\begin{eqnarray}
G_{\text{MF}}^{(2), T}(1,2)&=& \left(1 -w \cdot
G^{(2), T}_{\text{HS}}\left[\overline{\nu} + \varphi_{0} \right]
\right)^{-1} \cdot  G^{(2), T}_{\text{HS}}\left[\overline{\nu} +
\varphi_{0} \right]\left(1,2 \right)   \\
\label{zozo}
C_{\text{MF}}^{(2)}(1,2)&=&-G_{\text{MF}}^{(2), T \, \, -1}(1,2) = C_{\text{HS}}(1,2) + w(1,2) \; .
\end{eqnarray}
\end{subequations}
It follows then from eq.~(\ref{dens-K2}) that we have
\begin{equation}
\label{propa} G_{\varphi, \;\text{MF}}^{(2), T}(1,2)= \left(1 -w
\cdot G^{(2)}_{\text{HS}}\left[\overline{\nu} + \varphi_{0}
\right]\right)^{-1}
 \cdot w(1,2) \; .
\end{equation}

\subsection{\label{MF-CV}CV representation}
An analysis similar to that of Sec~\ref{MF-KSSHE} can be made in the CV representation.
The MF level of the CV field theory will be defined by
\begin{equation}
\Xi_{\text{MF}}\left[\nu \right]=\exp \left(- \mathcal{H}_{\text{CV}}\left[
\nu, \rho_{0}, \omega_{0} \right]  \right) \; ,
\end{equation}
where, for $\rho=\rho_{0}$ and $\omega=\omega_{0}$ the CV action is stationary, i.e.
\begin{equation}
\label{statio-CV}\left. \frac{\delta \; \mathcal{H}_{\text{CV}}\left[
\nu, \rho, \omega \right]}{\delta \rho}\right \vert_{(\rho_0,\omega_{0})}
=\left. \frac{\delta \; \mathcal{H}_{\text{CV}}\left[
\nu, \rho, \omega \right]}{\delta \omega}\right \vert_{(\rho_0,\omega_{0})}
=0 \; .
\end{equation}
Replacing the CV action by its expression~(\ref{actionCV}) in
eq.~(\ref{statio-CV}) yields a set of two coupled implicit
equations for $\rho_{0}$ and $\omega_{0}$:
\begin{eqnarray}
0&=& w(1,2)\rho_{0}(2) + i \omega_{0}(1) \nonumber \; , \\
0&=&  \rho_{0}(1) - \rho_{\text{HS}}\left[ \overline{\nu} -i\omega_{0} \right](1) \; .
\end{eqnarray}
If we define $\varphi_{0}= -i \omega_{0}$, then the two previous equations can be rewritten
\begin{eqnarray}
\varphi_{0}&=& w(1,2) \rho_{0}(2) \nonumber \; , \\
 \rho_{0}(1)&=& \rho_{\text{HS}}\left[ \overline{\nu} + \varphi_{0}\right](1) \; ,
\end{eqnarray}
which shows that, as expected, $\varphi_{0}$ coincides with the saddle point of the
KSSHE field theory (cf  Sec~\ref{MF-KSSHE}). Moreover a direct calculation will show that
\begin{equation}
\ln \Xi_{\text{MF}}\left[\nu \right]= -\mathcal{H}_{\text{K}}\left[\nu, \varphi_{0} \right]
= -\mathcal{H}_{\text{CV}}\left[\nu, \rho_{0},\omega_{0} \right] \; .
\end{equation}
Therefore the local density, the grand potential, the  free
energy, the  correlation and vertex functions coincide at the MF
level in the CV and KSSHE field theories.

\section{Loop expansion of the grand potential}
\label{loop}
A one loop expression for the grand potential $\ln
\Xi\left[\nu \right] $ and the  free energy $\beta
\mathcal{A}\left[\rho \right] $ has been obtained in
\cite{Cai-Mol} in the framework of KSSHE field theory. Here we
perform a two-loop expansion for $\ln \Xi$. The derivation will be
made both in the KSSHE and CV representations.

\subsection{\label{loop-KSSHE} KSSHE representation}
The loop expansion of the logarithm of the partition function of a
scalar field theory can be found in any standard text book, see
e.g. that of Zinn-Justin \cite{Zinn}. One proceeds as follows. A
small dimensionless parameter $\lambda$ is introduced and the loop
expansion is obtained by the set of transformations
\begin{eqnarray}
\varphi &=&\varphi_{0} + \lambda^{1/2} \chi \; ,\nonumber \\
\ln \Xi\left[ \nu \right] &=& \lambda \ln \left\lbrace
\mathcal{N}_{w}^{-1} \int \mathcal{D}\chi \;
\exp \left( - \frac{\mathcal{H}_{\text{K}}\left[\nu,\varphi \right]}{\lambda} \right)
\right\rbrace \; , \nonumber \\
 &=&\ln \Xi^{(0)}\left[ \nu \right] + \lambda \ln \Xi^{(1)}\left[ \nu \right]+
 \lambda^{2} \ln \Xi^{(2)}\left[ \nu \right] + \mathcal{O}(\lambda^{3}) \; ,
\end{eqnarray}
where $\varphi_{0}$ is the saddle point of the KSSHE action.  At
the end of the calculation one set $\lambda=1$. It follows from
the stationary condition~(\ref{statio-K}) that a functional Taylor
expansion of $\mathcal{H}_{\text{K}}\left[\nu,\varphi \right]$
about the saddle point has the form
\begin{eqnarray}
\label{loop-K}
\frac{\mathcal{H}_{\text{K}}\left[\nu,\varphi \right]}{\lambda} &=&
\frac{\mathcal{H}_{\text{K}}\left[\nu,\varphi_{0} \right]}{\lambda} +
\frac{1}{2} \left\langle \chi \vert \Delta_{\varphi_{0}}^{-1}\vert\chi \right\rangle \nonumber \\
&+& \sum_{n=3}^{\infty} \frac{\lambda^{\frac{n}{2} -1}}{n !}
\int_{\Omega} d1 \ldots dn \; \mathcal{H}^{(n)}_{\varphi_{0}}(1, \ldots,n)
\chi(1) \ldots \chi(n),
\end{eqnarray}
where we have adopted Zinn-Justin's notations \cite{Zinn}. In eq.~(\ref{loop-K})
 \begin{equation}
\label{propa-2}
\Delta_{\varphi_{0}}^{-1}(1,2)=
w^{-1}(1,2) -G_{\text{HS}}^{(2), \; T}\left[ \overline{\nu} +\varphi_{0}\right] (1,2)
\; .
\end{equation}
$\Delta_{\varphi_{0}}(1,2)$ is the free propagator of the theory, it coincides with
$G_{\varphi, \;\text{MF}}^{(2)}(1,2)$  as can be seen by comparing eqs.~(\ref{propa}) and~(\ref{propa-2}).
If $\nu$ is uniform then the system is homogeneous and $\Delta_{\varphi_{0}}(1,2)$ takes on a simple form in Fourier space, i.e.
\begin{equation}
 \widetilde{\Delta}_{\varphi_{0}}(q)=\frac{\widetilde{w}(q)}
{1- \widetilde{w}(q) \widetilde{G}_{\text{HS}}^{(2), \; T}\left[ \overline{\nu} +\varphi_{0}\right](q) } \; .
\end{equation}
Finally the kernels $\mathcal{H}^{(n)}_{\varphi_{0}}$ in the r.h.s. of eq.~(\ref{loop-K})
are given by
\begin{equation}
\label{kernel}
\mathcal{H}^{(n)}_{\varphi_{0}}(1, \ldots,n) \equiv -
G_{\text{HS}}^{(n), \; T}\left[ \overline{\nu} +\varphi_{0}\right] (1, \ldots,n) \; .
\end{equation}
The expression of $\ln \Xi^{(n)}\left[ \nu \right]$ in terms of
the propagator $\Delta_{\varphi_{0}}^{-1}(1,2)$ and the vertex interactions
$\mathcal{H}^{(n)}_{\varphi_{0}}$
is obtained by means of a cumulant expansion of  $\ln \Xi\left[ \nu \right]$ and by
making a repeated use of Wick's theorem.

Of course $ \Xi^{(0)}\left[ \nu \right]\equiv  \Xi_{\text{MF}}\left[ \nu \right]$.
At the one-loop order one finds \cite{Cai-Mol}
\begin{equation}
\label{Xi1-K} \Xi^{(1)}\left[ \nu \right]=
\frac{\mathcal{N}_{\Delta_{\varphi_{0}}}}{\mathcal{N}_{w}}=\frac{\int
\mathcal{D}\varphi \; \exp \left( -\frac{1}{2}\left \langle
\varphi \vert \Delta_{\varphi_{0}}\vert  \varphi \right \rangle
\right)}{\int \mathcal{D}\varphi \; \exp \left( -\frac{1}{2}\left
\langle \varphi \vert w \vert \varphi \right \rangle \right)  } \;
\end{equation}
For a homogeneous system the Gaussian integrals in the r.h.s. of
eq.~(\ref{Xi1-K}) can be performed explicitly (cf.
eqs.~(\ref{Gauss}) of Appendix A) and one has
\begin{equation}
\label{Xi1-K-bis}
\ln \Xi^{(1)}\left[ \nu \right]= -\frac{V}{2}
\int_{\mathbf{q}} \ln \left(1 - \widetilde{w}(q)   \widetilde{G}_{\text{HS}}^{(2), \; T}\left[ \overline{\nu}
+\varphi_{0} \right] (q)  \right) \; .
\end{equation}
As is shown in detail in \cite{Cai-Mol} the one-loop approximation
coincides with the well known RPA approximation of the theory of
liquids \cite{Hansen}.

\begin{figure}[!ht]
\begin{center}
\epsfig{figure=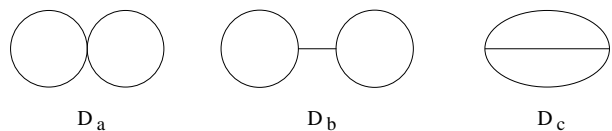,width=8cm,clip=}
\caption{\label{diag}
Diagrams which contribute to $ \ln \Xi^{(2)}\left[ \nu \right]$. $D_a$ and $D_c$ are irreducible while $D_b$ is reducible.}
\end{center}
\end{figure}

The second-loop order contribution $\ln \Xi^{(2)}$ to the  grand
potential has a complicated expression involving the sum of three
diagrams sketched in fig.~\ref{diag}
\begin{equation}
\label{Xi2-K}
\ln \Xi^{(2)}\left[ \nu \right]= D_a +D_b +D_c \; .
\end{equation}

More explicitly one has \cite{Zinn}
\begin{eqnarray}
\label{D}
D_a&=&
\frac{1}{8} \int_{\Omega} d1 \ldots d4 \;
\Delta_{\varphi_{0}}(1,2) \Delta_{\varphi_{0}}(3,4)
G_{\text{HS}}^{(4), \; T}\left[ \overline{\nu} +\varphi_{0} \right] (1,2,3,4) \nonumber \\
D_b&=&
\frac{1}{8} \int_{\Omega} d1 \ldots d3 \;  d1^{'} \ldots d3^{'}\;
\Delta_{\varphi_{0}}(1,2)  \Delta_{\varphi_{0}}(1^{'},2^{'}) \Delta_{\varphi_{0}}(3,3^{'})    \nonumber \\
& \times & G_{\text{HS}}^{(3), \; T}\left[ \overline{\nu} +\varphi_{0} \right] (1,2,3)
G_{\text{HS}}^{(3), \; T}\left[ \overline{\nu} +\varphi_{0} \right] (1^{'},2^{'},3^{'}) \nonumber \\
D_c&=&
\frac{1}{12} \int_{\Omega} d1 \ldots d3 \;  d1^{'} \ldots d3^{'}\;
\Delta_{\varphi_{0}}(1,1^{'}) \Delta_{\varphi_{0}}(2,2^{'}) \Delta_{\varphi_{0}}(3,3^{'}) \nonumber \\
& \times &
 G_{\text{HS}}^{(3), \; T}\left[ \overline{\nu} +\varphi_{0} \right] (1,2,3)
G_{\text{HS}}^{(3), \; T}\left[ \overline{\nu} +\varphi_{0} \right] (1^{'},2^{'},3^{'}) \; .
\end{eqnarray}
As they stand, the above relations are not particularly  useful
for practical applications (even in the homogeneous case) since
they involve the three and four body  correlation functions of the
reference HS fluid. We will introduce some reasonable
approximation in Sec.~\ref{free} to tackle with this horible
expression. Quite remarkably it has been shown recently that for a
symmetric mixture of charged hard spheres $\ln \Xi^{(2)}$ has a
much more simple expression which involves only the pair
correlation functions of the HS fluid as a consequence of local
charge neutrality \cite{Cai-JSP,Cai-Mol1}.

\subsection{\label{sec-loop-CV} CV representation}
In order to obtain a loop expansion of $\ln \Xi$ in the CV
representation we consider the following set of transformations
\begin{subequations}
\begin{eqnarray}
\label{aa}\rho &=&\rho_{0} + \lambda^{1/2} \delta \rho  \; , \\
\label{bb}\omega &=&\omega_{0} + \lambda^{1/2} \delta \omega \; , \\
\label{cc}\ln \Xi\left[ \nu \right] &=& \lambda \ln \left\lbrace  \int \mathcal{D}\delta\rho \;
\mathcal{D}\delta\omega \;
\exp \left( - \frac{\mathcal{H}_{\text{CV}}\left[\nu,\rho, \omega \right]}{\lambda} \right)
\right\rbrace \; ,
\end{eqnarray}
\end{subequations}
where $(\rho_{0},\omega_{0})$ is the saddle point of the CV
action. The form retained in eqs~(\ref{aa}) and~(\ref{bb}) is
imposed by the exact relation $\left\langle \omega
\right\rangle_{\text{CV}}= i w \star \left\langle \rho
\right\rangle_{\text{CV}} $ derived in Sec~\ref{Gnomega}. It
follows from the stationary condition~(\ref{statio-CV}) that the
functional Taylor expansion of
$\mathcal{H}_{\text{CV}}\left[\nu,\rho,\omega \right]$ about the
saddle point reads as
\begin{eqnarray}
\label{loop-CV}
\frac{\mathcal{H}_{\text{CV}}\left[\nu,\rho, \omega \right]}{\lambda}& =&
\frac{\mathcal{H}_{\text{CV}}\left[\nu,\rho_{0} ,\omega_{0} \right]}{\lambda}
- \frac{1}{2} \left\langle \delta \rho \vert  w \vert \delta \rho  \right\rangle
-i \left\langle \delta \omega \vert \delta \rho\right\rangle \nonumber \\
& -& \frac{1}{2}\left\langle \delta \omega \left\vert
G_{\text{HS}}^{(2), \;T}\left[\overline{\nu}-i \omega_{0} \right]
\right \vert\delta \omega \right\rangle
+\delta\mathcal{H}\left[\delta \omega \right] \; ,
\end{eqnarray}
where
\begin{equation}
\delta \mathcal{H}\left[\delta \omega \right]= - \sum_{n=3}^{\infty} \frac{ (-i)^{n}}{n !} \lambda^{\frac{n}{2} -1}
\int_{\Omega} d1 \ldots dn \; \mathcal{H}^{(n)}_{\varphi_{0}}(1, \ldots,n)
\delta \omega (1) \ldots \delta \omega (n) \; ,
\end{equation}
since, as $\varphi_{0}=-i \omega_{0}$ we have indeed
$G_{\text{HS}}^{(n), \;T}\left[\overline{\nu}-i \omega_{0}
\right]= - \mathcal{H}^{(n)}_{\varphi_{0}}$ where the kernel
$\mathcal{H}^{(n)}_{\varphi_{0}}$ is precisely that defined in
eq.~(\ref{kernel}).

We are thus led to define a two-fields Gaussian Hamiltonian
\begin{equation}
\mathcal{H}_{\text{G}}\left[\rho,\omega \right] \equiv
-\frac{1}{2}\left\langle \rho \vert w \vert \rho \right\rangle -i\left\langle\omega \vert \rho \right\rangle
+ \frac{1}{2}\left\langle \omega \vert
G^{(2), \; T}_{\text{HS}}\left[ \overline{\nu} -i \omega_{0} \right]
\vert \omega \right\rangle \; ,
\end{equation}
and Gaussian averages
\begin{equation}
\label{Gaverage}\left\langle \ldots \right\rangle_{\text{G}} \equiv
\mathcal{N}_{\text{G}}^{-1}
 \int \mathcal{D}\rho   \mathcal{D}\omega \; \ldots \exp\left(-
\mathcal{H}_{\text{G}}\left[\rho,\omega \right]  \right)
 \; ,
\end{equation}
where the normalization constant $\mathcal{N}_{\text{G}}$ is given by
\begin{equation}
\label{normaG}\mathcal{N}_{\text{G}}=\int \mathcal{D}\rho   \mathcal{D}\omega \;  \exp\left(-
\mathcal{H}_{\text{G}}\left[\rho,\omega \right]  \right) \: .
\end{equation}
Note that if $w>0$ (attractive case ) eq.~(\ref{Gaverage}) makes sense only if the integration
over the field variable $\omega$ is performed in first. With these notations in hands we can rewrite eq.~(\ref{cc}) as
\begin{equation}
\ln \Xi \left[ \nu \right]= \ln \Xi_{\text{MF}} \left[ \nu \right]
+ \lambda \ln \mathcal{N}_{\text{G}} + \lambda \ln
\left\langle  e^{\delta \mathcal{H}\left[ \omega \right]}\right\rangle_{\text{G}}  \; .
\end{equation}
A cumulant expansion of the last term in the r.h.s. yields the $\lambda$ expansion
\begin{equation}
\label{cumu}\ln \Xi = \ln \Xi_{\text{MF}}
+ \lambda \ln \overline{\Xi}^{(1)}+
\lambda \left\lbrace
\left\langle \delta \mathcal{H} \right\rangle_{\text{G}} +
\frac{1}{2}\left(  \left\langle \delta \mathcal{H}^{2} \right\rangle_{\text{G}}
-\left\langle \delta \mathcal{H} \right\rangle_{\text{G}}^{2}\right)
\right\rbrace +
 \mathcal{O}(\lambda^{3}) \; ,
\end{equation}
where
\begin{equation}
\overline{\Xi}^{(1)}\left[ \nu \right]=\mathcal{N}_{\text{G}} \; ,
\end{equation}
and it will be shown that the third term in the r.h.s. of
eq.~(\ref{cumu}) is of order $\mathcal{O}(\lambda^{2})$.

\subsubsection{One-loop approximation}
We want to prove that the CV and KSSHE one-loop approximations for $ \Xi$, respectively
$\overline{\Xi}^{(1)}$ and $\Xi^{(1)}$, coincide. Let us first rewrite the formula~(\ref{Xi1-K})
for $\Xi^{(1)}$ as the Gaussian average
\begin{equation}
\label{xx}\Xi^{(1)}\left[\nu \right]=
\left\langle
\exp \left( \frac{1}{2}
\left\langle \varphi \vert G_{\text{HS}}^{(2),\;T} \vert \varphi \right\rangle
\right) \right\rangle_{w} \; .
\end{equation}
Now we focus on $\overline{\Xi}^{(1)}\left[ \nu \right]=\mathcal{N}_{\text{G}}$.
Performing the (ordinary Gaussian) integration over $\omega$ first in eq.~(\ref{normaG}) we find
\begin{eqnarray}
\overline{\Xi}^{(1)}\left[\nu \right]&=&
\int \mathcal{D}\rho \;  \exp\left( \frac{1}{2}
\left\langle \rho \vert w \vert \rho  \right\rangle \right)
\int \mathcal{D}\omega\;  \exp\left(  i \left\langle \omega \vert \rho \right\rangle
 -\frac{1}{2}
\left\langle \omega \vert G_{\text{HS}}^{(2),\;T} \vert \omega
\right\rangle
 \right) \nonumber \\
 &=& \mathcal{N}_{\left[  G_{\text{HS}}^{(2),\;T}\right]^{-1}} \;
 \int \mathcal{D}\rho \;  \exp\left(- \frac{1}{2}
\left\langle \rho \vert \left[  G_{\text{HS}}^{(2),\;T}\right]^{-1}  - w \vert \rho  \right\rangle \right) \; .
\end{eqnarray}
Now we make use of the two following properties of Gaussian integrals
\begin{subequations}
\begin{eqnarray}
\label{propria}
 \mathcal{N}_{a}&=& 1/\mathcal{N}_{a^{-1}}  \\
\label{proprib} \left\langle \exp \left( \frac{1}{2}\left\langle
\varphi \vert b \vert \varphi  \right\rangle  \right)
\right\rangle_{a} &=&\left\langle \exp \left(
\frac{1}{2}\left\langle \varphi \vert a \vert \varphi
\right\rangle  \right) \right\rangle_{b} \; ,
\end{eqnarray}
\end{subequations}
to rewrite
\begin{eqnarray}
\overline{\Xi}^{(1)}\left[\nu \right]&=&\frac{
\int \mathcal{D}\rho \;  \exp\left(- \frac{1}{2}
\left\langle \rho \vert \left[  G_{\text{HS}}^{(2),\;T}\right]^{-1}  - w \vert \rho  \right\rangle \right)
}{\int \mathcal{D}\rho \;  \exp\left(- \frac{1}{2}
\left\langle \rho \vert \left[  G_{\text{HS}}^{(2),\;T}\right]^{-1}  \vert \rho  \right\rangle \right)} \nonumber \\
&=& \left\langle  \exp\left( \frac{1}{2}
\left\langle \rho \vert w  \vert \rho  \right\rangle \right)\right\rangle _{G_{\text{HS}}^{(2),\;T}} \;.
\end{eqnarray}
It follows then readily from eqs.~(\ref{xx}) and~(\ref{proprib})
that $\overline{\Xi}^{(1)}= \Xi^{(1)}$. The two
identities~(\ref{propria}) and ~(\ref{proprib}) are easy to show
in the homogeneous case and, in this case, there are an immediate
consequence of the properties of the Gaussian
integral~(\ref{Gauss}). In the general, non-homogeneous case, they
are derived in Appendix~B.

Let us first have a look at the correlations of the field $\rho$.
\subsubsection{Two-loop approximation}
At the two-loop level we have to compute the averages
$\left\langle \delta \mathcal{H}\left[\omega \right]
\right\rangle_{\text{G}}$ and $\left\langle \delta
\mathcal{H}\left[\omega \right]^{2} \right\rangle_{\text{G}}$. In
order to do so we have first to generalize Wick's theorem to this
special kind of Gaussian average defined in eq.~(\ref{Gaverage}).

Let us first have a look at the correlations of the field $\rho$. One has from eq.~(\ref{Gaverage})
\begin{eqnarray}
\left\langle \rho(1) \ldots \rho(n) \right\rangle_{\text{G}} &=&
\frac{
 \int \mathcal{D}\rho   \mathcal{D}\omega \; \rho(1) \ldots \rho(n) \exp\left(-
\mathcal{H}_{\text{G}}\left[\rho,\omega \right]  \right)}
{\int \mathcal{D}\rho \mathcal{D}\omega \; \exp\left(-
\mathcal{H}_{\text{G}}\left[\rho,\omega \right]  \right)
} \nonumber \\
&=& \frac{
 \int \mathcal{D}\rho  \; \rho(1) \ldots \rho(n) \exp\left(
-\frac{1}{2} \left\langle \rho \left \vert \left[  G_{\text{HS}}^{(2),\;T}\right]^{-1}  -w  \right \vert \rho
\right\rangle \right)}
{\int \mathcal{D}\rho \; \exp\left(- \frac{1}{2} \left\langle \rho \left \vert \left[  G_{\text{HS}}^{(2),\;T}
\right]^{-1}  -w \right \vert \rho  \right\rangle
  \right)
}
\nonumber\; ,
\end{eqnarray}
where we have performed the (ordinary) Gaussian integral on the
field $\omega$ first. We are thus led to the ordinary Gaussian
average (provided that $\left[ G_{\text{HS}}^{(2),\;T}\right]^{-1}
-w$ is positive definite, which we assume), therefore the usual
Wick's theorem is applied which yields
\begin{eqnarray}
\left\langle \rho(1) \right\rangle_{\text{G}}^{T} &=& 0 \nonumber \\
\left\langle \rho(1)\rho(2) \right\rangle_{\text{G}}^{T} &=& \left( \left[ G_{\text{HS}}^{(2),\;T}
\right]^{-1} -w\right)^{-1}=G_{\text{MF}}^{(2), \; T}(1,2) \nonumber \\
\left\langle \rho(1) \ldots \rho(n) \right\rangle_{\text{G}}^{T}&=& 0 \; (\text{for } n\geq 3) \; .
\end{eqnarray}
We take now advantage of the general relations~(\ref{dens-CV}) between the truncated
correlations of $\rho$ and $\omega$ in the CV formalism to get
\begin{eqnarray}
\left\langle \omega(1) \right\rangle_{\text{G}}^{T} &=& 0 \nonumber \\
\left\langle \omega(1)\omega(2) \right\rangle_{\text{G}}^{T} &=& -w(1,1^{'}) w(2,2^{'})
G_{\text{MF}}^{(2), \; T}(1^{'},2^{'})
=-\Delta_{\varphi_{0}}(1,2) \nonumber \\
\left\langle \omega(1) \ldots \omega(n) \right\rangle_{\text{G}}^{T}&=& 0 \; (\text{for } n\geq 3) \; ,
\end{eqnarray}
from which Wick's theorem follows
\begin{eqnarray}
\label{WickG}
\left\langle \omega(1) \ldots \omega(n) \right\rangle_{\text{G}}&=& 0 \; (n \text{ odd }) \; , \nonumber \\
\left\langle \omega(1) \ldots \omega(n) \right\rangle_{\text{G}}&=&
(-1)^{n}\sum_{\text{pairs}}\prod \Delta_{\varphi_{0}}(i_{1},i_{2})\; (n \text{ even }) \; ,
\end{eqnarray}
which was an expected result thanks to the correspondence $\varphi_{K} \leftrightarrow i \omega_{CV}$.
We have now at our disposal all the tools to compute
\begin{eqnarray}
\label{morceau1}\left\langle \delta \mathcal{H}\left[\omega \right] \right\rangle_{\text{G}}&=&
\frac{(-i)^{4}}{4!} \lambda \int d1 \ldots d4 \;
G_{\text{HS}}^{(4)}\left[\overline{\nu}+\varphi_{0} \right]
(1,\ldots 4) \nonumber \\
&\times&
\left\langle \omega(1) \ldots \omega(4) \right\rangle_{\text{G}} +
\mathcal{O}(\lambda^{2}) \nonumber \\
&=&\frac{\lambda}{8} \int d1 \ldots d4
\;G_{\text{HS}}^{(4)}\left[\overline{\nu}+\varphi_{0} \right]
(1,\ldots 4) \nonumber \\
&\times & \Delta_{\varphi_{0}}(1,2)\Delta_{\varphi_{0}}(3,4) +\mathcal{O}(\lambda^{2}) \; ,
\end{eqnarray}
where we have made use of Wick's theorem~(\ref{WickG}).
We note that $\left\langle \delta \mathcal{H}\left[\omega \right] \right\rangle_{\text{G}}^{2}$
will not contribute to $\ln \Xi$ at the two-loop order and it remains to compute
$\left\langle \delta \mathcal{H}^{2}\left[\omega \right] \right\rangle_{\text{G}}$. One finds
\begin{eqnarray}
\label{morceau2}
\left\langle \delta \mathcal{H}^{2}\left[\omega \right] \right\rangle_{\text{G}}&=&
\frac{\lambda}{(3!)^{2}}
\int d1 \ldots d3^{'}
G_{\text{HS}}^{(3)}\left[\overline{\nu}+\varphi_{0} \right] (1,2,3)
G_{\text{HS}}^{(3)}\left[\overline{\nu}+\varphi_{0} \right] (1^{'},2^{'},3^{'}) \nonumber \\
&\times&\left\langle \omega(1) \ldots \omega(3^{'}) \right\rangle_{\text{G}} +
\mathcal{O}(\lambda^{2}) \nonumber \\
&=&
\lambda \int d1 \ldots d3^{'}
G_{\text{HS}}^{(3)}\left[\overline{\nu}+\varphi_{0} \right] (1,2,3)
G_{\text{HS}}^{(3)}\left[\overline{\nu}+\varphi_{0} \right] (1^{'},2^{'},3^{'})\nonumber \\
&\times &\left\lbrace
\frac{1}{4}\Delta_{\varphi_{0}}(1,2) \Delta_{\varphi_{0}}(1^{'},2^{'})\Delta_{\varphi_{0}}(3,3^{'})\right.   \nonumber \\
&+& \left. \frac{1}{6}\Delta_{\varphi_{0}}(1,1^{'}) \Delta_{\varphi_{0}}(2,2^{'})\Delta_{\varphi_{0}}(3,3^{'})\right\rbrace  + \mathcal{O}(\lambda^{2}) \; ,
\end{eqnarray}
where once again Wick's theorem~(\ref{WickG}) was used. Gathering the intermediate results~(\ref{morceau1})
and~(\ref{morceau2}) one has finally, after inspection
\begin{equation}
\lambda \left\lbrace
\left\langle \delta \mathcal{H} \right\rangle_{\text{G}} +
\frac{1}{2}\left(  \left\langle \delta \mathcal{H}^{2} \right\rangle_{\text{G}}
-\left\langle \delta \mathcal{H} \right\rangle_{\text{G}}^{2}\right)
\right\rbrace = \lambda^{2} \;\ln \overline{\Xi}^{(2)}\left[\nu \right] +
\mathcal{O}(\lambda^{3})
\end{equation}
with
\begin{equation}
\overline{\Xi}^{(2)}\left[\nu \right]=\Xi^{(2)}\left[\nu \right] \; .
\end{equation}
We have thus shown that the one and two-loop expressions for $\ln
\Xi\left[\nu \right]$ coincide in the KSSHE and CV
representations. This coincidence is likely to be  true at all
orders in the loop-expansion.

\section{The pressure and the free energy of the homogeneous fluid}
\label{free}
\subsection{The pressure}
In this section we restrict ourselves to the homogeneous case,
therefore $\ln \Xi\left[\nu \right]=V \beta P\left(\nu \right)$,
where $P$ denotes the pressure and $\beta \mathcal{A}\left[\rho
\right]= V \beta f(\rho)$ where $f$ is the Helmoltz free energy
per unit volume. The two-loop expression that we derived for $P$
in Sec.~\ref{loop} is too much complicated to be of any use in
practical calculations since it involves the $3$ and $4$ density
correlation functions of the HS fluid which are unknown, whereas
$G_{\text{HS}}^{(2), \; T}$ is known approximatively, for instance
in the Percus-Yevick (PY) approximation \cite{Hansen}. A simple
but coherent approximations for $G_{\text{HS}}^{(3), \; T}$ and
$G_{\text{HS}}^{(4), \; T}$ will be proposed now.

Recall first that it follows from their definitions \cite{Stell1,Stell2} (see e.g. eqs~(\ref{defcorre}))
that the $G_{\text{HS}}^{(n), \; T}\left[\nu \right] $ satisfy to the following relations
\begin{equation}
\label{h1}\frac{\delta }{\delta \nu(n+1)}
G_{\text{HS}}^{(n), \; T}\left[\nu \right](1, \ldots,n)=G_{\text{HS}}^{(n+1), \; T}\left[\nu \right](1, \ldots,n,n+1) \; .
\end{equation}
For a homogeneous system (in which case  $\nu$ is a constant) one infers from this equation that
\begin{equation}
\label{h2}\int_{\Omega}d1 \ldots dn \; G_{\text{HS}}^{(n+1), \; T}\left[\nu \right](1, \ldots,n,n+1)=
\frac{\partial^{n}}{\partial \nu^{n}}\rho_{\text{HS}}\left(\nu \right)\equiv
\rho_{\text{HS}}^{(n)}\left(\nu \right)
\; ,
\end{equation}
where $\rho_{\text{HS}}\left(\nu \right)$ is the number density of
hard spheres at the chemical potential $\nu$.

In the rest of the section we will adopt the following approximation
\begin{equation}
\label{hyp}
G_{\text{HS}}^{(n), \; T}\left[\nu \right](1, \ldots,n)=
\rho_{\text{HS}}^{(n+1)}\left(\nu \right)\delta(n,1) \ldots \delta(2,1) \; \text{for n} \geq 3 \; .
\end{equation}
Note that this hypothesis is coherent with the exact
relations~(\ref{h1}) and~(\ref{h2}). However,  for $n=2$, we adopt for
$G_{\text{HS}}^{(2), \; T}\left[\nu \right](1,2)$  some  known approximation of the theory of liquids (PY
approximation for instance). The free propagator has then  a non trivial
expression. In Fourier space it reads
\begin{equation}
\widetilde{\Delta}_{\varphi_{0}}(q)=\frac{\widetilde{w}(q)}{
1-\widetilde{w}(q) \widetilde{G}_{\text{HS}}^{(2), \; T}\left[\overline{\nu}+\varphi_{0}\right](q) },
\end{equation}
where
\begin{equation}
\widetilde{G}_{\text{HS}}^{(2), \; T}(q)=
\int d^{3}x \;  e^{-i q x} G_{\text{HS}}^{(2), \; T}(x)
\end{equation}
is the Fourier transform of $G_{\text{HS}}^{(2), \; T}\left[\nu
\right](x)$. The set of approximations that we have just discussed
is reasonable as long as the range of the KSSHE field correlation
functions is (much) larger than than the range of the HS density
correlation functions. This would be true if $w$ is the long range
pair interaction or, in general, in the vicinity of the critical
point.

With the hypothesis~(\ref{hyp}) it is easy to obtain the two-loop
order approximation for the pressure. One finds
\begin{eqnarray}
\label{press}\beta P (\nu) &=&\beta P^{(0)}(\nu) + \lambda \beta P^{(1)}(\nu)+
\lambda^{2} \beta P^{(2)}(\nu)+\mathcal{O}(\lambda^{3}) \; ,\nonumber \\
\beta P^{(0)}(\nu)&=&\beta P_{\text{MF}}(\nu)=P_{\text{HS}}(\overline{\nu}+\varphi_{0})-\frac{\varphi_{0}^{2}}{
2 \widetilde{w}(0)} \; ,\nonumber \\
\beta P^{(1)}(\nu)&=&-\frac{1}{2} \int_{\mathbf{q}} \ln\left(
1- \widetilde{w}(q)\widetilde{G}_{\text{HS}}^{(2), \; T}\left[
\overline{\nu}+\varphi_{0}
\right](q) \right) \; ,\nonumber \\
\beta P^{(2)}(\nu)&=&\frac{\rho^{(3)}_{0}}{8}\Delta_{\varphi_{0}}^{2}(0)+
\frac{\left[ \rho^{(2)}_{0}\right] ^{2}}{8} \widetilde{\Delta}_{\varphi_{0}}(0) \Delta_{\varphi_{0}}^{2}(0) \nonumber \\
&+& \frac{\left[ \rho^{(2)}_{0}\right] ^{2}}{12}\int d^{3}x \; \Delta_{\varphi_{0}}^{3}(x) \; ,
\end{eqnarray}
where $\rho^{(n)}_{0} \equiv \rho_{\text{HS}}^{(n)}\left( \overline{\nu}+\varphi_{0}\right)$ and
$\int_{\mathbf{q}}\equiv \int d^{3}q/(2 \pi)^{3}$.

\subsection{The free energy}
In order to compute the free energy $f(\rho)$ at the two-loop
order we need the expression of the density only at the one-loop
order (this a well known properties of Legendre transform
\cite{Zinn} that will be checked explicitely further on). We thus
define
\begin{eqnarray}
\label{end1}\rho(\nu)&=&\frac{\partial }{\partial \nu} \beta P=\rho_0 + \Delta \rho(\nu) \nonumber \\
\Delta \rho(\nu)&=&\lambda \rho^{(1)}\left(\nu \right) + \mathcal{O}(\lambda^{2})\;,
\end{eqnarray}
where $\rho_0\equiv \rho_{\text{HS}}\left( \overline{\nu}+\varphi_{0}\right)$ and
\begin{eqnarray}
\label{www}\rho^{(1)}\left(\nu \right)&=&\frac{\partial}{\partial \nu}\beta P^{(1)}(\nu)
=\frac{1}{2}\int_{\mathbf{q}} \frac{\widetilde{w}(q)}{
1-\widetilde{w}(q) \widetilde{G}_{\text{HS}}^{(2), \; T}(q)} \;
\frac{\partial}{\partial \nu} \widetilde{G}_{\text{HS}}^{(2), \; T}(q) \nonumber \\
&=& \frac{\rho_{0}^{(2)}}{2} \Delta_{\varphi_{0}}(0)\left( 1+\frac{\partial \varphi_{0}}{\partial \nu}
\right) \; ,
\end{eqnarray}
where, in order to derive the last line, we made use of eqs.~(\ref{h1}) and~(\ref{hyp}). In order to obtain the
expression of $\partial \varphi_{0}/\partial \nu$ one derives the homogeneous stationary condition~(\ref{statio2-K-hom})
with respect to $\nu$ which gives us
\begin{equation}
1+\frac{\partial \varphi_{0}}{\partial \nu}=\frac{1}{1 - \rho_{0}^{(1)}\widetilde{w}(0)} \; .
\end{equation}
Introducing this result in the last line of eq.~(\ref{www}) we
find finally that
\begin{equation}
\label{ro1}
\rho^{(1)}\left(\nu \right)=\frac{\rho_{0}^{(2)}}{2} \frac{\Delta_{\varphi_{0}}(0)}{1 - \rho_{0}^{(1)}\widetilde{w}(0)} \; .
\end{equation}
We are now in position to compute the free energy
\begin{eqnarray}
\beta f (\rho) &=& \rho \nu - \beta P(\nu) \nonumber \\
&=& \rho \nu -\beta P^{(0)}(\nu)  - \lambda \beta P^{(1)}(\nu)-
\lambda^{2} \beta P^{(2)}(\nu)+\mathcal{O}(\lambda^{3}) \; .
\end{eqnarray}
Our task is now to reexpress the r.h.s. as a function of
$\rho=\rho_0 + \Delta \rho(\nu)$ which will be done along the same
lines that those used in \cite{Cai-JSP}. We shall compute
separately three contributions to $\beta f$.
\\\\
\noindent \textit{i) Computation of} $X\equiv\rho \nu -\beta P^{(0)}(\nu) $
\\\\
We first note that $X$ can be rewritten as
\begin{equation}
X=\rho_0 \nu -\beta P^{(0)}(\nu) + \nu \Delta \rho =\beta f^{(0)}(\rho_0)+
\nu \Delta \rho \; ,
\end{equation}
where $f^{(0)}(\rho_0)\equiv f_{\text{MF}}(\rho_0) $ is the MF
free energy of the homogeneous system at the  density $\rho_0$,
i.e. (cf. eq.~(\ref{MF-A}))
\begin{equation}
\label{12}
\beta f^{(0)}(\rho_0)=\beta f_{\text{HS}}(\rho_0) -\frac{1}{2} \rho_0 \widetilde{w}^{2}(0)+
\frac{1}{2} \rho_0 w(0) \; .
\end{equation}
In order to relate $f^{(0)}(\rho_0)$ to $f(\rho]$ we perform a second order Taylor expansion
\begin{equation}
\label{13}\beta f^{(0)}(\rho =\rho_0 + \Delta \rho )=
\beta f^{(0)}(\rho_0) + \nu \Delta \rho -\frac{1}{2} \widetilde{C}_{MF}(0)
\left[ \Delta \rho \right]^{2} + \mathcal{O}(\lambda^{3}) \; ,
\end{equation}
where the Fourier transform at $q=0$ of the MF two-body vertex
function is computed from eq.~(\ref{zozo}) with the result
\begin{equation}
\label{15} \widetilde{C}_{MF}(0)=-1/\widetilde{G}_{HS}(0)
+\widetilde{w}(0)=-\frac{1-\rho_{0}^{(1)}
\widetilde{w}(0)}{\rho_{0}^{(1)}} \; .
\end{equation}
Combining the intermediate results~(\ref{12}),~(\ref{13}), and~(\ref{15}) one has finally
\begin{eqnarray}
\label{X} X & = &\beta f^{(0)}(\rho) + \frac{1}{2}\label{14}\widetilde{C}_{MF}(0)
 \left[ \Delta \rho \right]^{2} + \mathcal{O}(\lambda^{3}) \nonumber \\
 &=&\beta f^{(0)}(\rho) -\frac{\lambda^{2}}{8} \Delta_{\varphi_{0}}^{2}(0)
 \frac{\left[ \rho_{0}^{(2)}\right] ^{2}}{\rho_{0}^{(1)}\left(
 1 - \rho_{0}^{(1)} \widetilde{w}(0)\right) } + \mathcal{O}(\lambda^{3}) \; .
\end{eqnarray}
We note that, as claimed at the beginning of the paragraph, the knowledge of $\Delta \rho$
at the one-loop order is sufficient for our purpose.
\\\\
\noindent \textit{ii) Computation of} $\beta P^{(1)}(\nu) $
\\\\
As can be inferred from eq.~(\ref{press}) $\beta P^{(1)}(\nu)$ may
be seen as a function of the sole variable $\rho_0\equiv
\rho_{HS}(\overline{\nu}+\varphi_{0})$ since for the HS reference
system there is a one to one correspondence between densities and
chemical potentials -at least in the fluid phase, away from the
liquid/solid transition-. Therefore $\beta P^{(1)}(\nu)
\equiv\beta P^{(1)}(\rho_0) $, with some abusive notations. One
thus have
\begin{eqnarray}
\beta P^{(1)}(\rho=\rho_0+ \Delta \rho)&=&\beta P^{(1)}(\rho_0) + \frac{\partial \nu}{\partial\rho_0 }
\frac{\partial \beta P^{(1)}(\nu)}{\partial \nu} \Delta \rho +
\mathcal{O}(\lambda) \; , \nonumber \\
&=&\beta P^{(1)}(\rho_0) + \frac{\partial \nu}{\partial\rho_0 }
\frac{\left[\Delta \rho \right]^{2}}{\lambda} +  \mathcal{O}(\lambda) \; ,
\end{eqnarray}
where the last line follows from eqs.~(\ref{end1}) and~(\ref{www}).  At this point we notice that
\begin{equation}
\frac{\partial \rho_0}{\partial \nu}=
\widetilde{G}_{\text{MF}}^{T}(q=0)= \frac{\rho_0^{(1)}}{1 - \rho_0^{(1)}
\widetilde{w}(0)}=1/\frac{\partial \nu}{\partial\rho_0 } \; ,
\end{equation}
which allows us to write finally
\begin{subequations}
\begin{eqnarray}
\beta P^{(1)}(\rho_0)&=& \beta P^{(1)}(\rho) -\frac{\lambda}{4}
\Delta^{2}_{\varphi_{0}}(0)
\frac{\left[\rho_0^{(2)}\right]^{2}}{\rho_0^{(1)} \left( 1-
\rho_0^{(1)}\widetilde{w}(0) \right) } \\
\beta P^{(1)}(\rho)&=& - \frac{1}{2}\int_q \ln \left(1 -
 \widetilde{w}\left( q\right)  \widetilde{G}_{\text{HS}, \; \rho}^{T}\left( q\right)
 \right)  \; ,
\end{eqnarray}
\end{subequations}
where, in the second equality, the subscript emphasizes that the
truncated HS pair correlation function has to be computed at the
density $\rho$.

The computation of $f(\rho)$ is now completed and gathering the
intermediate results one has finally, setting $\lambda=1$
\begin{eqnarray}
\label{f2l}
\beta f(\rho)&=& \beta f_{\text{HS}}(\rho) -\frac{\widetilde{w}\left( 0\right)}{2}
\rho^{2} \nonumber \\
&+&\frac{1}{2} \int_{\mathbf{q}} \left\lbrace \ln \left(1 -
 \widetilde{w}\left( q\right)  \widetilde{G}_{\text{HS}, \; \rho}^{T}\left( q\right)
 \right)  + \rho \widetilde{w}\left( q\right)\right\rbrace  \nonumber \\
 &-&\frac{\rho^{(3)}_{0}}{8}\Delta_{\rho}^{2}(0)+
\frac{1}{8}  \Delta_{\rho}^{2}(0)\frac{\left[ \rho^{(2)}_{0}\right]^{2} }{\rho^{(1)}_{0}}
-  \frac{\left[ \rho^{(2)}_{0}\right] ^{2}}{12}\int d^{3}x \;
\Delta_{\rho}^{3}(x) \; .
\end{eqnarray}
Some comments on this result are in order.

i) The last line in eq.~(\ref{f2l}) gathers the two-loop
contributions of $\beta f(\rho)$. At this order of the loop
expansion it is thus legitimate to replace $\Delta_{\varphi_0}$ by
$\Delta_{\rho}$ where the propagator is evaluated at the density
$\rho$ rather than $\rho_0$ since $\rho -
\rho_0=\mathcal{O}(\lambda)$. More explicitly
\begin{equation}
\widetilde{\Delta}_{\rho}(q)=\frac{\widetilde{w}\left( q\right)}{1 -\widetilde{G}_{\text{HS}, \; \rho}^{T}
\left( q\right)\widetilde{w}\left( q\right) } \; .
\end{equation}

ii) Similarly the derivatives $\rho^{(n)}_{0}$ which enter the
last line of eq.~(\ref{f2l}) can be evaluated at the density
$\rho$ rather than $\rho_0$. One can thus write, taking advantage
of the one to one correspondence between the HS densities and
chemical potentials
\begin{eqnarray}
\rho^{(1)}_{0}&=&\frac{1}{\nu_{\text{HS}}^{(1)}\left( \rho\right) } \nonumber \\
\rho^{(2)}_{0}&=&\frac{\partial\rho^{(1)}_{0} }{\partial \nu} =\frac{-\nu_{\text{HS}}^{(2)}(\rho)}
{\left[\nu_{\text{HS}}^{(1)}(\rho)\right]^{3}} \nonumber \\
\rho^{(3)}_{0}&=&\frac{\partial\rho^{(2)}_{0} }{\partial \nu} =
\frac{3 \left[\nu_{\text{HS}}^{(2)}(\rho)\right]^{2} -
\nu_{\text{HS}}^{(3)}(\rho)\nu_{\text{HS}}^{(1)}(\rho)
}{\left[\nu_{\text{HS}}^{(1)}(\rho)\right]^{5}} \; ,
\end{eqnarray}
where $\nu_{\text{HS}}^{(n)}(\rho)$ denotes the $n$th derivative
of the HS chemical potential with respect to the density (it can
be computed in the framework of the PY or Carnahan-Starling
approximations for instance \cite{Hansen}).

iii) It must be pointed out that, quite unexpectedly, the
reducible diagram $D_b$ has not be cancelled by the Legendre
transform. Usually, in statistical field theory it is the case
(cf. \cite{Zinn}). The reason is that  the chemical potential
$\nu$ is not the field conjugate to the order parameter
$m=\left\langle \varphi \right\rangle_{K}$ of the KSSHE field
theory. However one of us  have shown elsewhere
\cite{Cai-JSP,Cai-Mol1} that for the symmetric mixtures of charged
hard spheres only irreducible diagrams contribute to $\beta
f^{(2)}$.

iv) All the quantities which enter eq.~(\ref{f2l}) can be computed
numerically (for instance in the PY approximation). We plan to test
the validity of our approximation in future work.

\section{Conclusion}
Using the CV method we reconsider the basic relations of
statistical field theory of simple fluids that follow from this
approach. In contrary to the KSSHE theory \cite{Cai-Mol} the
corresponding CV action depends on two scalar fields - field
$\rho$ connected to the number density of particles and field
$\omega$ conjugate to $\rho$. Explicit expressions that allow to
relate between themselves the correlation functions defined in
different versions of the theory are derived.

For a one-component continuous model of fluids, consisting of hard
spheres interacting through the short-range pair potential, we
calculated the grand partition function in a systematic way for both versions of
statistical field theory using the loop expansion technique. As it
was expected, in all the orders of loop expansion considered, both
versions of the theory produced indeed the same analytical
results. The expressions for the pressure as well as for the free
energy are derived in a two-loop approximation that is the next
step in comparison with the results obtained recently by one
of us \cite{Cai-Mol} within the KSSHE theory. In fact this is a
new type of approximation which we plan to test in our future
work.

In contrast to the usual statistical field theory \cite{Zinn}, our results demonstrate that for the case of simple fluids the reducible diagram is not cancelled by the Legendre transform. It is due to the fact that our field theory is a non-standard in the sense that the coupling between the internal field $\varphi$ and the external field $\nu$ is non-linear yielding  new expressions for $\beta f(\rho)$.

From our analysis of the CV and KSSHE transformations we can also
conclude that the former has some important advantages which could
be very useful for more complicate models of fluids. In
particular, it is valid for an arbitrary pair potential (including
a pair interaction $w(1,2)$ which does not possess an inverse) and
is easily generalized for the case of n-body interparticle
interactions with $n>2$.

\section*{Acknowledgments}
This work was made in the framework of the cooperation project
between the CNRS and the NASU (ref.~ CNRS 17110).
\appendix
\label{appendixA}
\section{Functional measures and integrals}
In this appendix we give some details on functional measures and
integrals. Let us consider a real scalar field
$\varphi(\mathbf{x})$ defined in a cube $\mathcal{C}_3$ of side
$L$ and volume $V=L^3$. We assume periodic boundary conditions,
i.e. we restrict ourselves to fields which can be written as a
Fourier series,
\begin{equation}
\varphi(\mathbf{x})=\frac{1}{L^3} \; \sum_{\mathbf{q} \in \Lambda}
\widetilde{\varphi}_{\mathbf{q}} \; e^{i \mathbf{q}\mathbf{x}} \;
,
\end{equation}
where $\Lambda = (2 \pi/L)\;  \Z^3$ is the reciprocal cubic
lattice. The reality of $\varphi$ implies that, for $\mathbf{q}\ne
0$ $\widetilde{\varphi}_{-\mathbf{q}} =
\widetilde{\varphi}_{\mathbf{q}}^{\star}$, where the star means
complex conjugation. Following Wegner \cite{Wegner} we define the
normalized functional measure $\mathcal{D}\varphi$ as
\begin{subequations}
\label{dphi}
\begin{eqnarray}
\mathcal{D} \varphi & \equiv & \prod_{\mathbf{q} \in \Lambda}
\frac{d \widetilde{\varphi}_{\mathbf{q}} }
{\sqrt{2 \pi  V}} \\
d \widetilde{\varphi}_{\mathbf{q}} d
\widetilde{\varphi}_{-\mathbf{q}} & = & 2 \;
d\Re{\widetilde{\varphi}_{\mathbf{q}}} \;
d\Im{\widetilde{\varphi}_{\mathbf{q}}} \text{ for } \mathbf{q} \ne
0 \; .
\end{eqnarray}
\end{subequations}
Eq.\ (\ref{dphi}) can  be conveniently rewritten as
\begin{equation}
\label{dphi_bis} \mathcal{D} \varphi= \frac{d \varphi_{0}}{\sqrt{2
\pi  V}} \prod_{\mathbf{q} \in \Lambda^{\star}}
\frac{d\Re{\widetilde{\varphi}_{\mathbf{q}}} \;
d\Im{\widetilde{\varphi}_{\mathbf{q}}} }{\pi V} \; ,
\end{equation}
where the sum in the r.h.s runs over only the half $\Lambda^{*}$ of
all the vectors of the reciprocal lattice
$\Lambda$ (for instance those with $q_x \geq 0$). With the definition~(\ref{dphi})
we have
\begin{eqnarray}
\label{norma}
\mathcal{N}_{w}&=& \int \mathcal{D} \varphi \;  \exp \left( -\frac{1}{2}
\left\langle \varphi \vert w^{-1} \vert \varphi \right\rangle \right) \nonumber \\
&=&
\exp \left(\frac{1}{2 } \sum_{{\mathbf{q}} \in \Lambda}  \ln \widetilde{w}(q) \right)  \nonumber \\
&\xrightarrow{L \to \infty}&\exp \left( \frac{V}{2}\;
\int_{\mathbf{q}} \ln \widetilde{w}(q) \right) \; ,
\end{eqnarray}
where $w(1,2)$ is positive and satisfies $w(1,2)=w(2,1)\equiv
w(x_{12})$. It is worth noting that in this case we have trivially
$\mathcal{N}_{w^{-1}}=1/\mathcal{N}_{w}$. More generally and with
the same hypothesis we have the useful identity
\begin{eqnarray}
\label{Gauss}
\left\langle \exp \left( (i)\left\langle \varphi \vert  \omega \right\rangle \right)
\right\rangle_{w}
&\equiv&
\mathcal{N}_{w}^{-1} \int \mathcal{D} \varphi \;
\exp \left( -\frac{1}{2}
\left\langle \varphi \vert w^{-1} \vert \varphi \right\rangle + (i)
\left\langle \varphi \vert  \omega \right\rangle \right) \nonumber \\
&=&
\exp \left(
 +(-) \frac{1}{2}\left\langle \omega \vert w \vert \omega \right\rangle
\right) \; ,
\end{eqnarray}
where $\omega$ is a real scalar field.

We define now the "functional delta" distribution $\delta_{\mathcal{F}}\left[ \lambda \right] $  as \cite{Orland}
\begin{equation}
\label{deltaF}
\delta_{\mathcal{F}}\left[ \lambda\right] \equiv
 \int \mathcal{D} \omega \; \exp \left(i \left\langle \omega
 \vert \lambda  \right\rangle \right) \;,
\end{equation}
where both $\omega$ and $\lambda$ are \textit{real} scalar fields defined on $\mathcal{C}$. Since
\begin{equation}
\left\langle \omega
 \vert \lambda  \right\rangle=
\int_{\mathcal{C}} d^3\mathbf{x} \; \omega(\mathbf{x}) \lambda (\mathbf{x})=\frac{1}{V}\sum_{\mathbf{q} \in \Lambda}
 \widetilde{\omega}_{-\mathbf{q}}\widetilde{\lambda}_{\mathbf{q}} \; ,
\end{equation}
it follows from eq.~(\ref{dphi}) that we have more explicitly
\begin{equation}
\label{deltaF-bis} \delta_{\mathcal{F}}\left[ \lambda\right]=
\sqrt{2 \pi V} \; \delta \left( \widetilde{\lambda}_0 \right)  \;
\prod_{\mathbf{q} \in \Lambda^{\star}}\left[ \pi V \; \delta
\left( \Re{\widetilde{\lambda}_{\mathbf{q}}} \right)  \; \delta
\left( \Im{\widetilde{\lambda}_{\mathbf{q}}} \right) \right] \; .
\end{equation}
Therefore
\begin{equation}
\int \mathcal{D} \lambda \;\delta_{\mathcal{F}}\left[ \lambda\right]=1 \;,
\end{equation}
and, more generally
\begin{equation}
\int \mathcal{D} \lambda \;
\mathcal{F}\left[\lambda \right]
\delta_{\mathcal{F}}\left[ \lambda
- \lambda _0\right]=\mathcal{F}\left[\lambda_0 \right]
 \;,
\end{equation}
where  $\mathcal{F}\left[\lambda \right] $ is some arbitrary functional
of the field $\lambda(x)$.
\label{appendixB}
\section{Two useful identities}
We prove here two identities used in Sec.~\ref{loop}.
Firstly let $K(\mathbf{x},\mathbf{y})$ be a definite positive operator assumed to be symmetric.
The variables $\mathbf{x}$ and $\mathbf{y}$ take their values in the cube $\mathcal{C}_{3}$ with periodic boundary conditions. We define (provided that it exists)
\begin{equation}
\mathcal{N}_K=\int \mathcal{D} \varphi \; \exp\left( -\frac{1}{2}
\left\langle \varphi \vert K^{-1} \vert \varphi \right\rangle
\right) \; ,
\end{equation}
and we want to prove that $\mathcal{N}_{K^{-1}}=1/\mathcal{N}_K$.
If $K(\mathbf{x},\mathbf{y})=K(\mathbf{x}-\mathbf{y})=K(\mathbf{y}-\mathbf{x})$, i.e. in the homogeneous case, it is
trivial as pointed out in Appendix A. In the general case we still
have \cite{Zinn}
\begin{equation}
\int \mathcal{D} \varphi \; \exp\left( -\frac{1}{2}
\left\langle \varphi \vert K^{-1} \vert \varphi \right\rangle + i\left\langle J \vert
\varphi \right\rangle \right)=\mathcal{N}_K \exp\left( -\frac{1}{2}
\left\langle J \vert K \vert J \right\rangle
\right) \; ,
\end{equation}
where $J$ is some real scalar field. Therefore
\begin{eqnarray}
\mathcal{N}_{K^{-1}}&=& \int \mathcal{D} J \; \exp\left( -\frac{1}{2}
\left\langle J \vert  K \vert J \right\rangle
\right) \nonumber \\
&=&\frac{1}{\mathcal{N}_{K}}
\int \mathcal{D} J \mathcal{D} \varphi \;
\exp \left(  -\frac{1}{2} \left\langle \varphi \vert K^{-1} \vert \varphi \right\rangle + i\left\langle J \vert
\varphi \right\rangle\right) \nonumber \\
&=&\frac{1}{\mathcal{N}_{K}} \int\mathcal{D} \varphi \;
\exp \left( -\frac{1}{2}  \left\langle \varphi \vert K^{-1} \vert \varphi \right\rangle\right)
\delta_{\mathcal{F}}\left[ \varphi\right] \nonumber \\
&=&\frac{1}{\mathcal{N}_{K}} \; ,
\end{eqnarray}
where we have made use of the representation~(\ref{deltaF}) of $\delta_{\mathcal{F}}\left[ \varphi\right]$.

Secondly let us consider two definite positive symmetric operators $a(\mathbf{x},\mathbf{y})$ and $b(\mathbf{x},\mathbf{y})$.
We define (provided that these objects exist)
\begin{equation}
\label{gt}\left\langle \exp\left( \frac{1}{2}
\left\langle \varphi \vert a \vert \varphi  \right\rangle
\right)  \right\rangle_{b}=\frac{
\int\mathcal{D} \varphi \; \exp\left( - \frac{1}{2}\left\langle \varphi \vert b^{-1}
\vert \varphi  \right\rangle + \frac{1}{2}\left\langle \varphi \vert a \vert \varphi  \right\rangle\right)
}{\int\mathcal{D} \varphi \; \exp\left( - \frac{1}{2}\left\langle \varphi \vert b^{-1} \vert \varphi
\right\rangle \right)} \; ,
\end{equation}
and we want to prove that
\begin{equation}
\label{tt}\left\langle \exp\left( \frac{1}{2}
\left\langle \varphi \vert a \vert \varphi  \right\rangle
\right)  \right\rangle_{b}=\left\langle \exp\left( \frac{1}{2}
\left\langle \varphi \vert b \vert \varphi  \right\rangle
\right)  \right\rangle_{a} \; .
\end{equation}
Note that these expressions are well defined only if $(a-b^{-1})>0$ and
$(b-a^{-1})>0$, i.e. $ab<1$.
If $a(b)(\mathbf{x},\mathbf{y})=a(b)(\mathbf{x}-\mathbf{y})=a(b)(\mathbf{y}-\mathbf{x})$ (i.e. in the homogeneous case) the result is a trivial
consequence of eq.~(\ref{Gauss}).

In the general case one proceeds as follows. Since $a>0$ and $b>0$ then one can define the
operators $a^{\pm 1/2}$ and $b^{\pm 1/2}$ which are all symmetric and positive definite.
Let us perform the change of variable
\begin{equation}
\varphi(1)=b^{1/2}(1,1^{'}) a^{-1/2}(1^{'},1^{''}) \psi(1^{''})
\end{equation}
in eq.~(\ref{gt}). The functional Jacobian of the transform $J= \det
\delta \varphi(1)/ \delta \psi (2)=\det b^{1/2}(1,1^{'}) a^{-1/2}(1^{'},2)$ is a non zero constant
which can be picked out of the integral. Moreover it is easy to show that
\begin{equation}
\left\langle \varphi \vert b^{-1} \vert \varphi \right\rangle=
\left\langle \psi \vert a^{-1} \vert \psi \right\rangle,   \qquad
\left\langle \varphi \vert a \vert \varphi \right\rangle =\left\langle \psi \vert b \vert \psi \right\rangle
\end{equation}
therefore one has
\begin{eqnarray}
\left\langle \exp\left( \frac{1}{2}
\left\langle \varphi \vert a \vert \varphi  \right\rangle
\right)  \right\rangle_{b}&=&\frac{
J \int \mathcal{D} \psi \exp\left( -\frac{1}{2}
\left\langle \psi \vert a^{-1}\vert \psi \right\rangle + \frac{1}{2}
\left\langle \psi \vert b \vert \psi \right\rangle
\right) }{J \int \mathcal{D} \psi \exp\left( -\frac{1}{2}
\left\langle \psi \vert a^{-1}\vert \psi \right\rangle
\right)} \nonumber \\
&=&\left\langle \exp\left( \frac{1}{2}
\left\langle \varphi \vert b \vert \varphi  \right\rangle
\right)  \right\rangle_{a} \; .
\end{eqnarray}

\end{document}